# Using x-ray diffraction to identify precipitates in transition metal doped semiconductors

Shengqiang Zhou, K. Potzger, G. Talut, J. von Borany, W. Skorupa, M. Helm, and J. Fassbender

Institute of Ion Beam Physics and Materials Research, Forschungszentrum Dresden-Rossendorf, P.O. Box 510119, 01314 Dresden, Germany

In the past decade, room temperature ferromagnetism was often observed in transition metal doped semiconductors, which were claimed as diluted magnetic semiconductors (DMS). Nowadays intensive activities are devoted to clarify wether the observed ferromagnetism stems from carrier mediated magnetic impurities, ferromagnetic precipitates, or spinodal decomposition. In this paper, we have correlated the structural and magnetic properties of transition metal doped ZnO, TiO$_2$, and Si, prepared by ion implantation. Crystalline precipitates, i.e., transition metal (Co, Ni) and Mn-silicide nanocrystals, are responsible for the magnetism. Additionally due to their orientation nature with respect to the host, these nanocrystals in some cases are not detectable by conventional x-ray diffraction (XRD). This nature results in the pitfall of using XRD to exclude magnetic precipitates in DMS materials.

## 1. INTRODUCTION

Formation of ferromagnetic precipitates is a major obstacle in the fabrication of diluted magnetic semiconductors (DMSs).[1,2,3,4] These ferromagnetic precipitates can dominate the observed ferromagnetism,[5,6,7] and alter the magnetotransport properties.[8] Therefore, to prove the existence or nonexistence of precipitates is highly desired to understand the ferromagnetism in DMSs. X-ray diffraction (XRD) is often employed to exclude the crystalline precipitates.[9,10,11] In this paper, we will address the difficulty of using XRD to detect nanocrystals embedded inside semiconductors.
The intensity of diffracted x rays by a plane (*hkl*) is given by

$$I_{hkl} \propto I_0 V |F_{hkl}|^2 / v^2$$

where $I_0$ is the intensity of incident beam, $V$ the volume of the crystal, $v$ the volume of the unit cell, and $F_{hkl}$ the structure factor.[12] In order to improve the detection limit, for a specific sample, one should increase $I_0$, i.e., using a synchrotron-radiation x-ray source, and measure a diffraction plane with a large $F_{hkl}$. Moreover the diffraction from the substrate is always much stronger than from precipitates. Thus, reducing the influence of the substrate by, e.g., grazing incidence, also improves the visibility of precipitates. The structure factor $F_{hkl}$ is tabulated by the International Center for Diffraction Data (ICDD). In general, a crystalline material has several diffraction planes with large structure factors, while other diffraction planes exhibit structure factors smaller by one or more order of magnitude. For instance, the relative structure factors of hcp-Co(10$\underline{1}$1), (0002), and (10$\underline{1}$2) are 100, 60, and 1, respectively (PDF 5-727).

Additionally, the formation of nanoscale magnetic precipitates can be indirectly identified from their magnetic properties. If magnetic nanoparticles are sufficiently small, above the blocking temperature $T_B$, thermal fluctuations dominate and no preferred magnetization direction can be defined. Phenomenologically there are two characteristic features in the temperature dependent magnetization of a nanoparticle system. One is the irreversibility of the magnetization in a small applied field (e.g., 50 Oe) after zero field cooling and field cooling (ZFC/FC).[13] The other is the drastic drop of the coercivity and of the remanence at a temperature close to or above $T_B$.[14]

By correlating the structural and magnetic properties, we have analyzed the formation of transition metal (Co, Ni) and Mn-silicide nanocrystals in semiconductors. We attempt to generalize the difficulties in probing nanocrystals of small amount using XRD.

## 2. EXPERIMENTS

Commercial ZnO(0001), rutile $TiO_2$(110), and Si(001) single crystals were implanted with transition metal ions. Three kinds of samples as shown in Table 1 will be presented. All samples were investigated using XRD and superconducting quantum interference device (SQUID) magnetometry. Synchrotron-radiation XRD (SR-XRD) was performed at the Rossendorf beamline (BM20) at the ESRF with an x-ray wavelength of 0.154 nm. Laboratory-equipped XRD (Lab-XRD) was performed using a Siemens D5005 with an x-ray wavelength of 0.154 056 nm.

## 3. RESULTS

### 3.1 Crystallographically oriented precipitates

Figures 1a,1b show $2\vartheta$–$\vartheta$ scans of Co implanted ZnO by Lab-XRD and SR-XRD, respectively. Concerning the formation of crystalline precipitates, both techniques reveal the same trend. At a low fluence (0.8×10$^{16}$ cm$^{-2}$), no crystalline precipitates could be detected, while from a fluence of 4×10$^{16}$ cm$^{-2}$ the hcp-Co(0002) peak appears, and grows with increasing fluence. Figure 1c shows the $\phi$-scans of hcp-Co(10$\underline{1}$1) and ZnO(10$\underline{1}$1). Both scans show a sixfold symmetry at the same azimuthal position. hcp-Co is crystallographically oriented inside ZnO matrix. The orientation relationship is hcp-Co(0001)[1$\underline{1}$00] // ZnO(0001)[1$\underline{1}$00]. Figure 1d shows the $2\vartheta$–$\vartheta$ scan for hcp-Co(10$\underline{1}$1) in a skew geometry at one of the azimuthal positions. In skew geometry, the incident and the diffracted waves have the same angles to the surface, while the sample is tilted with respect to its surface normal. By this configuration, a noncoplanar, its surface normal does not lie in the plane defined by the incident and the diffracted waves, can be measured.[15] By this approach, we also can confirm that the peaks in Figs. 1a,1b are from hcp-Co(0002), not fcc-Co(111).

Fig 1.

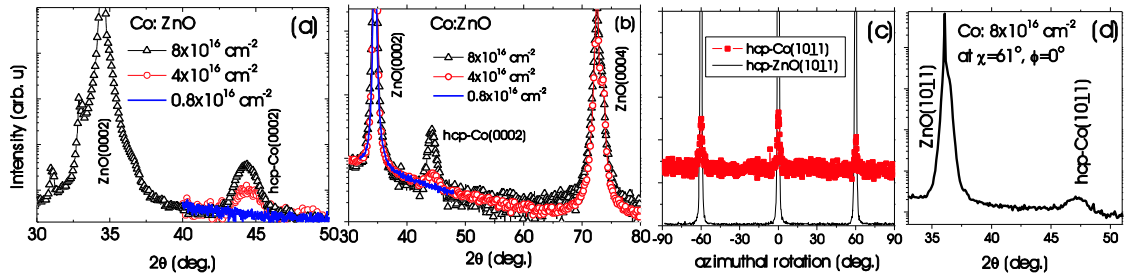

Structural properties of Co implanted ZnO: (a) $2\vartheta-\vartheta$ scan of ZnO(0002) and hcp-Co(0002) by Lab-XRD; (b) $2\vartheta-\vartheta$ scan of ZnO(0002)(0004) and hcp-Co(0002) by SR-XRD; (c) $\phi$-scan of Co(10$\bar{1}$1) and ZnO(10$\bar{1}$1); (d) $2\vartheta-\vartheta$ scan of ZnO(10$\bar{1}$1) and hcp-Co(10$\bar{1}$1) by SR-XRD in a skew geometry.

Similar to the case of Co in ZnO, Ni nanocrystals are also crystallographically oriented following the relationship of Ni(111)[11$\bar{2}$] ∥ ZnO(0001)[1$\bar{1}$00].[16] Consequently Co and Ni precipitates are rather easy to be detected even by Lab-XRD.[6] In contrast, bcc-Fe is not crystallographically oriented inside ZnO matrix due to the fourfold symmetry of bcc-Fe. These crystalline Fe precipitates are only detected by SR-XRD given the same implantation fluence of $4\times10^{16}$ cm$^{-2}$.[5]

Figure 2 shows the ZFC/FC magnetization curves of Co implanted ZnO. The field is perpendicular to the sample surface. An irreversible behavior is observed in ZFC/FC curves. The broad peak in ZFC curves is due to the size distribution of Co NCs. The inset of Fig. 2 shows the hysteresis loops. At 300 K, both coercivity and remanence drop to around zero.

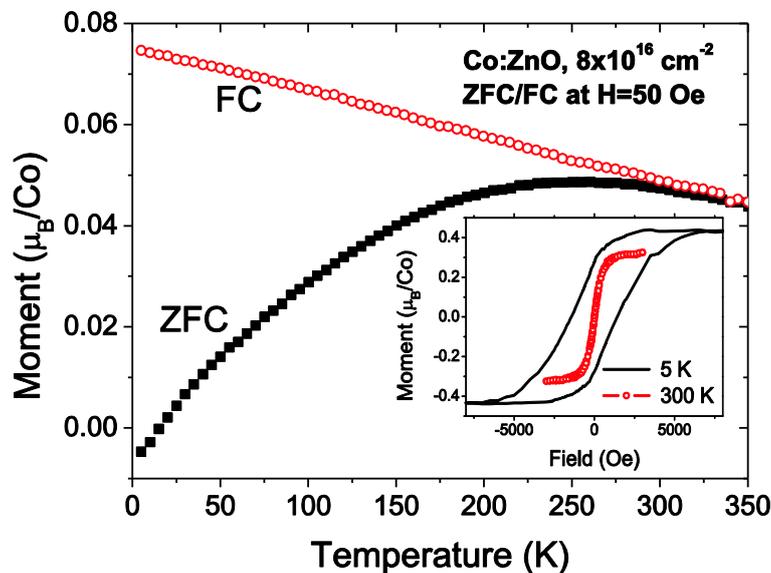

Fig 2.

ZFC/FC magnetization under an applied field of 50 Oe. Inset: Hysteresis loops measured at 5 and 300 K.

## 3.2 Misoriented precipitates

Figure 3a shows $2\vartheta-\vartheta$ scan of Co implanted $TiO_2$ by SR-XRD. One broad shoulder besides $TiO_2(220)$ is observed at the position of hcp-Co(10$\bar{1}$2). Note that the diffraction plane (10$\bar{1}$2) has a much smaller structure factor than (10$\bar{1}$1), which means that (10$\bar{1}$1) must not be parallel with the sample surface. hcp-Co, if there is, is mis oriented inside $TiO_2$. With the assumption of Co(10$\bar{1}$2) // $TiO_2$(110), Co(10$\bar{1}$1) should be tilted by 49.82° or 19.78° from sample surface. Figure 3b shows $\phi$-scans of Co(10$\bar{1}$1). As expected, two-fold-symmetric Co(10$\bar{1}$1) is observed. It is reasonable since rutile $TiO_2$(110) is also two-fold-symmetric viewed along the perpendicular direction. Figure 3c shows $2\vartheta-\vartheta$ scan of Co(10$\bar{1}$1). The Co(10$\bar{1}$1) plane is parallel with $TiO_2$(210). Figure 3d shows ZFC/FC curves measured with an applied field of 50 Oe. Like hcp-Co in ZnO (Fig. 2), the ZFC/FC curves show temperature irreversibility. The inset shows the temperature dependent coercivity and remanence. Both values decrease quickly with increasing temperatures, and drop to zero at around 100 K, just above $T_B$, the maximum in ZFC curve. Note that $T_B$ is smaller than that shown in Fig. 2. This is due to different Co implantation fluence. A larger fluence results in larger Co precipitates, consequently a larger $T_B$.

Fig3.

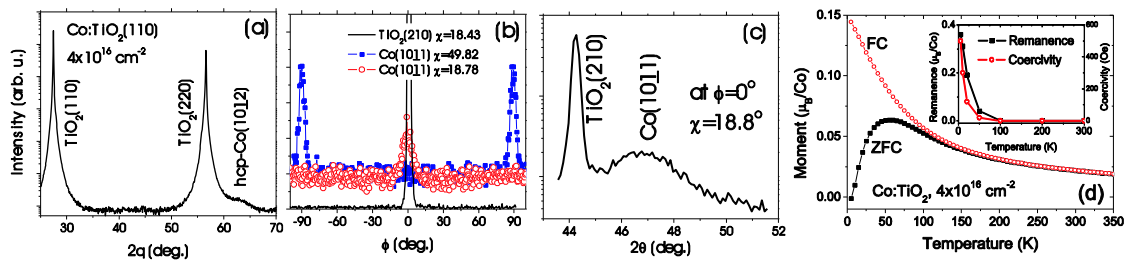

Structural and magnetic properties of Co implanted $TiO_2$: (a) $2\vartheta-\vartheta$ scan of $TiO_2$(110)(220) by SR-XRD; (b) $\phi$-scan of Co(10$\bar{1}$1) and $TiO_2$(210); (c) $2\vartheta-\vartheta$ scan of $TiO_2$(210) and hcp-Co(10$\bar{1}$1) by SR-XRD in a skew geometry; (d) ZFC/FC magnetization under an applied field of 50 Oe. Inset: temperature dependent coercivity and remanence.

## 3.3 Randomly oriented precipitates

For Mn-implanted Si(001), rapid thermal annealing (RTA) was performed at 1073 K for 5 min in $N_2$ flow. Lab-XRD reveals both as-implanted and RTA samples free of crystalline precipitates. Moreover, even at SR-XRD in a symmetric beam geometry, one fails to detect any Mn-silicides [Fig. 4]. Therefore, a grazing incidence geometry was used during the measurement, where the incident x-ray beam is aligned at a small (here 0.4°) angle to the sample surface. The diffraction peaks at around 42° and 46.3° cannot be attributed to the Si substrate, but to $MnSi_{1.7}$. Detailed structural and magnetic properties of Mn implanted Si have been reported in Ref. 7.

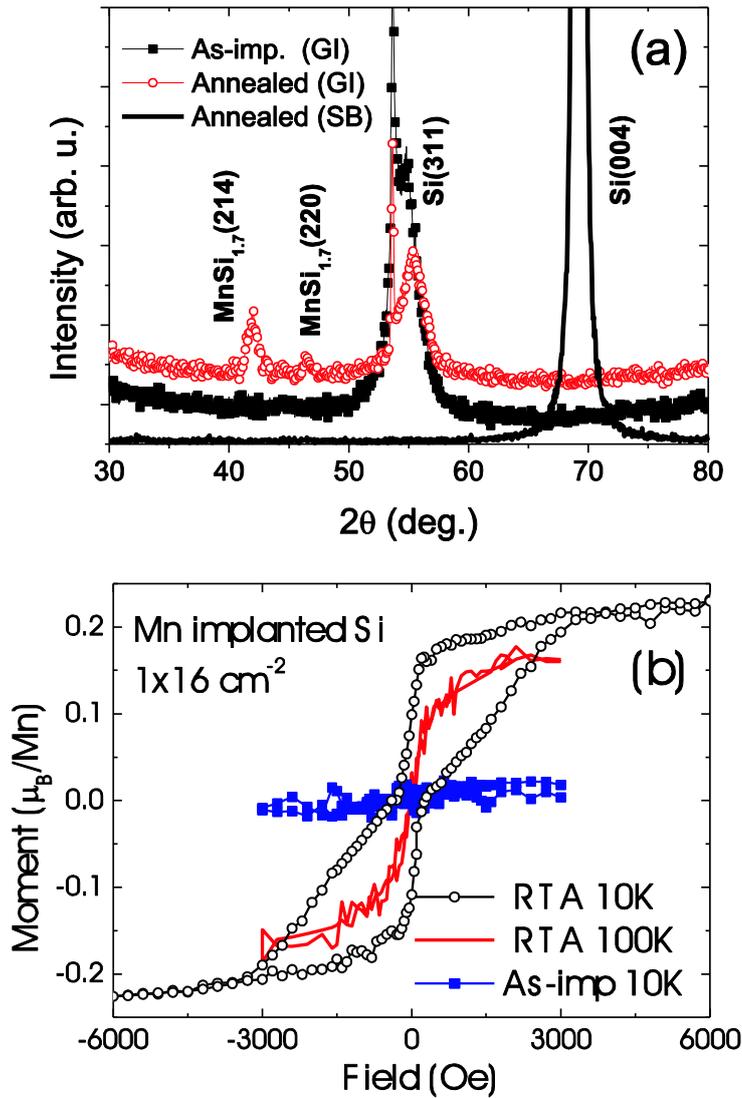

Fig 4.

SR-XRD patterns of Mn implanted Si. GI: grazing incidence, and SB: symmetric beam path.

## 4. CONCLUSIONS

In conclusion, we have shown the formation of magnetic precipitates in transition metal doped semiconductors, and addressed the difficulty to detect these nanocrystals. The orientation between nanocrystals and substrates can be divided into three categories as shown in Fig. 5. Figure 5a shows the case of crystallographic orientation, e.g., the case of Ni and Co in ZnO, with one diffraction plane with large structure factor parallel with the sample surface. Figure 5b shows misorientation (e.g., hcp-Co in $TiO_2$), where the precipitates are also crystallographically oriented inside substrates; however, the diffraction planes with large structure factors are not parallel with sample surface. The last case is random orientation as shown in Fig. 5c. Obviously, in a $2\vartheta-\vartheta$ scan, the most often used method in phase identification by XRD, crystalline precipitates are easy to be detected in the first case since all precipitates contribute to the diffraction intensity. For the latter two cases, the techniques presented here must be applied to identify the precipitates since the conventional technique is not sufficiently sensitive.

Fig 5.

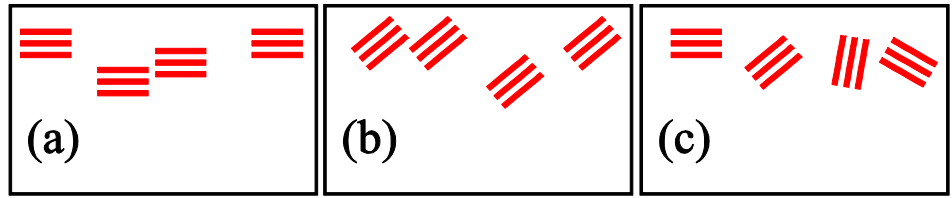

Schematic orientation of nanocrystalline precipitates with respect to the substrate.

## Tables

Table I. Sample preparation: implantation parameters.

| Substrate | Dopant | Energy (keV) | Temperature (K) | Fluence (×10$^{16}$ cm$^{-2}$) |
|---|---|---|---|---|
| ZnO | Co | 180 | 623 | 0.8–8 |
| TiO$_2$ | Co | 180 | 623 | 4 |
| Si | Mn | 300 | 573 | 1 |

1. R. Seshadri, Curr. Opin. Solid State Mater. Sci. **9**, 1 (2005).
2. C. Liu, F. Yun, and H. Morkoc, J. Mater. Sci.: Mater. Electron. **16**, 555 (2005).
3. R. Janisch, P. Gopal, and N. A. Spaldin, J. Phys.: Condens. Matter **17**, R657 (2005).
4. T. Dietl, J. Phys.: Condens. Matter **19**, 165204 (2007).
5. K. Potzger, S. Q. Zhou, H. Reuther, A. Mücklich, F. Eichhorn, N. Schell, W. Skorupa, M. Helm, J. Fassbender, T. Herrmannsdorfer et al., Appl. Phys. Lett. **88**, 052508 (2006).
6. S. Zhou, K. Potzger, G. Zhang, F. Eichhorn, W. Skorupa, M. Helm, and J. Fassbender, J. Appl. Phys. **100**, 114304 (2006). |
7. S. Zhou, K. Potzger, G. Zhang, A. Mücklich, F. Eichhorn, N. Schell, R. Grotzschel, B. Schmidt, W. Skorupa, M. Helm et al., Phys. Rev. B **75**, 085203 (2007). |
8. M. Jamet, A. Barski, T. Devillers, V. Poydenot, R. Dujardin, P. Bayle-Guillemaud, J. Rothman, E. Bellet-Amalric, A. Marty, J. Cibert et al., Nat. Mater. **5**, 653 (2006).
9. Y. Matsumoto, M. Murakami, T. Shono, T. Hasegawa, T. Fukumura, M. Kawasaki, P. Ahmet, T. Chikyow, S. Koshihara, and H. Koinuma, Science **291**, 854 (2001).
10. N. H. Hong, W. Prellier, J. Sakai, and A. Hassini, Appl. Phys. Lett. **84**, 2850 (2004).
11. X. Liu, F. Lin, L. Sun, W. Cheng, X. Ma, and W. Shi, Appl. Phys. Lett. **88**, 062508 (2006).
12. C. G. Darwin, Philos. Mag. **27**, 315 (1914). |
13. M. Respaud, J. M. Broto, H. Rakoto, A. R. Fert, L. Thomas, B. Barbara, M. Verelst, E. Snoeck, P. Lecante, A. Mosset et al., Phys. Rev. B **57**, 2925 (1998). |

14. S. R. Shinde, S. B. Ogale, J. S. Higgins, H. Zheng, A. J. Millis, V. N. Kulkarni, R. Ramesh, R. L. Greene, and T. Venkatesan, Phys. Rev. Lett. **92**, 166601 (2004).
15. V. M. Kaganer, O. Brandt, A. Trampert, and K. H. Ploog, Phys. Rev. B **72**, 045423 (2005). [
16. S. Zhou, K. Potzger, J. von Borany, R. Grötschel, W. Skorupa, M. Helm, and J. Fassbender, Phys. Rev. B **77**, 035209 (2008).